\documentclass[twoside,11pt]{article}

\usepackage[accepted]{melba}
\usepackage[final]{changes}
\usepackage{bm}

\definechangesauthor[name=ivan, color=blue]{ID} 
\definechangesauthor[name=richard, color=blue]{RM}

\usepackage{amsmath,amsfonts,comment}

\usepackage{xcolor}
\usepackage{graphicx}
\usepackage{subcaption}
\melbaid{2024:010}  
\doi{https://doi.org/10.59275/j.melba.2024-7189}
\melbaauthors{Diaz, Geiger and McKinley}  
\volume{2}
\firstpageno{834}  
\melbayear{2024}  
\datesubmitted{03/2023}  
\datepublished{05/2024}  
\ShortHeadings{ SO(3)-steerable convolutions for pose-robust 3D segmentation}{Diaz, Geiger and McKinley}

\title{Leveraging SO(3)-steerable convolutions for pose-robust semantic segmentation in 3D medical data}

\author{\name Ivan Diaz \email ivan.diaz@insel.ch \\  
\addr Support Center for Advanced Neuroimaging (SCAN), University Institute of Diagnostic and Interventional Neuroradiology, University of Bern, Inselspital, Bern University Hospital, Bern, Switzerland
    \AND
    \name Mario Geiger \email mgeiger@nvidia.com  \\
    \addr NVIDIA, Santa Clara, USA
 \AND Richard Iain McKinley\email richard.mckinley@insel.ch  \\
    \addr Support Center for Advanced Neuroimaging (SCAN), University Institute of Diagnostic and Interventional Neuroradiology, University of Bern, Inselspital, Bern University Hospital, Bern, Switzerland
}
\begin{document}

\maketitle

\begin{abstract}
Convolutional neural networks (CNNs) allow for parameter sharing and translational equivariance by using convolutional kernels in their linear layers. By restricting these kernels to be SO(3)-steerable, CNNs can further improve parameter sharing \deleted{and equivariance}. These \added{rotationally-}equivariant convolutional layers have several advantages over standard convolutional layers, including increased robustness to unseen poses, smaller network size, and improved sample efficiency.
Despite this, most segmentation networks used in medical image analysis continue to rely on standard convolutional kernels. In this paper, we present a new family of segmentation networks that use equivariant voxel convolutions based on spherical harmonics. These networks are robust to data poses not seen during training, and do not require rotation-based data augmentation during training. In addition, we demonstrate improved segmentation performance in MRI brain tumor and healthy brain structure segmentation tasks, with enhanced robustness to reduced amounts of training data and improved parameter efficiency.

Code to reproduce our results, and to implement the equivariant segmentation networks for other tasks is available at~\url{http://github.com/SCAN-NRAD/e3nn_Unet}.
\end{abstract}


\section{Introduction}

A \emph{symmetry} of an object is a transformation of that object which leaves certain properties of that object unchanged.  In the context of medical image segmentation, there are a number of obvious symmetries which apply to volumetric images and their voxel-level labels: namely translation, rotation, and (depending on the labels used) reflection across the body's left-right axis of symmetry.  In most cases patients are placed in an expected orientation within the scanner (with fetal imaging being a notable exception to this assumption), and deviations from the mean patient placement are typically moderate (typically up to 20 degrees). Nonetheless, given a small data set, the patient orientations seen may not be representative of the full range of poses seen in clinical practice.

An equivariant function is one where symmetries applied to an input lead to corresponding transformations of the output. The most prominent example of equivariance in deep learning is the translation-equivariance of the convolution operation. Equivariance should be contrasted to mere \emph{invariance}, where a symmetry applied to an input leads to no change in a function's output. The output of a segmentation model should not be invariant to symmetries of its input, but rather equivariant.  Equivariance enables increased parameter sharing and enforces strong priors which can prevent overfitting and improve sample efficiency.  

There have been numerous attempts to define convolutional feature extractors equivariant to rotational (and reflection) symmetry in three dimensional space.  Since voxelized data (in contrast to point cloud data) only admits rotations through 90 degrees, an obvious place to start is the symmetries of the cube.  Group equivariant convolutional networks (G-CNNs) \citep{cohen2016group}, in the context of 3D imaging, operate by applying transformed versions of a kernel according to a \emph{finite} symmetry group $\mathcal{G}$.  This gives rise to an extra fiber/channel dimension with size $| \mathcal{G} |$ (24 in total if only considering orientation-preserving symmetries of the cube, or 48 if considering all symmetries), which permute under symmetries of the input.  This results in an explosion in the number of convolutional operations and in  the dimension of feature maps.  \emph{G-pooling} can be used to combat this explosion, by selecting the fiber channel which maximizes activation at each voxel.  This reduces memory usage but comes at the cost of reducing the expressivity of the layer, potentially impacting performance \citep{cohen2016group}.

Steerable convolutions with full rotational equivariance to infinite symmetry groups in three dimensions were first developed for point cloud data \citep{thomas_tensor_2018}, and have subsequently been adapted to operate voxel convolutions on data lying in regular 3D grids  \citep{weiler_3d}. These convolutional layers have the benefit, over G-CNN layers, of being equivariant to any 3D rotation, rather than a discrete group of rotations: in particular, the rotations likely to arise as a result of patient placement in a scanner. They are also more efficient in terms of convolution operations and memory usage.  The \texttt{e3nn} \citep{e3nn, geiger2022e3nn} pytorch library provides a flexible framework for building  SE(3) equivariant (translation, rotation) as well as E(3) (translation, rotation and reflection) networks for both point cloud and voxel data, by providing implementations of SO(3) and O(3) steerable kernels\footnote{E(3) refers to the Euclidean group in 3 dimensions, SE(3) the special Euclidean group in 3 dimensions, O(3) the orthogonal group in 3 dimensions and SO(3) the special orthogonal group in three dimensions. }.  These kernels operate on irreducible representations (irreps), which provide a general description of equivariant features: any finite representation which transforms according to the group action of SO(3)/O(3) can be expressed as a direct sum of irreps.

Methods based on steerable filters have long been used in biomedical image analysis, but \emph{learnable} steerable filters have not received much attention, despite the promised benefits. This may be because of perceived computational overheads, or the lack of available code for building such networks.  Our goal in this paper is to show that the benefits of equivariance, sample efficiency and parameter efficiency can be made available in biomedical image analysis without sacrificing performance.  To this end, we make use of equivariant max-pooling and normalization layers as well as equivariant voxel convolutions and use them to formulate a standard 3D Unet architecture in which each layer, and therefore the whole network, is equivariant. 
\added{It is important to note that our discussion on equivariance is primarily rooted in a theoretical framework where operations are defined within the continuous domain. Here, equivariant max-pooling is
conceptualized as selecting the vector with the maximum amplitude in a vicinity around
each point, which, in theory, ensures strict equivariance due to its consistent behavior under
transformations. However, the necessity of discretization for practical implementation introduces an element
of approximation to this idealized equivariance. Our reference to equivariance pertains to this theoretical, continuous domain model, where we assumed the impact of discretization on equivariance would be minimal. It's in the transition from continuous to discrete domain that approximate equivariance emerges, primarily due to the discretization inherent in rasterization.}

Our primary hypothesis is as follows: end-to-end rotation equivariant networks provide robustness to data orientations unseen during training without loss of performance on in-sample test data, beyond the robustness gained by using rotational data augmentation \citep{mei2021learning}.  We further hypothesize that equivariant networks have better sample efficiency than traditional Unets.

\section{Building an equivariant segmentation network}

In this paper we focus on leveraging the benefit of convolutional kernels which exhibit SE(3)-equivariance. It is easy to extend our work to E(3) but we leave this for future work. Note that while most of the layers we define are fully equivariant to translations and 90 degree rotations, rotations through intermediate angles cannot be shown to be theoretically equivariant, owing to the nature of voxelized data.  In addition, pooling (as used in typical Unet architectures) is not even translation equivariant\citep{xu2021group} and further exacerbates the effects of discretization by changing the spatial regions over which features are aggregated. We aim therefore for equivariant primitive operations, and conduct experiments to validate that this theoretical equivariance also yields practically useful approximate robustness to rotations.  In common with other authors, we use the term \emph{equivariant} network/model to refer to our proposed network, since it is built from primitives with improved equivariance with respect to standard architectures.  \deleted{This is not intended to communicate theoretical equivariance beyond 90 degree rotations.}

The Unet architecture \citep{ronneberger2015u} consists of an encoding path and decoding path with multiple levels: on each level there are multiple convolutions and nonlinearities, followed by either a pooling or upsampling operation. To achieve SE(3) equivariance in a neural network it is necessary that each of these operations be equivariant. We use the steerable 3D convolution and gated nonlinearity described in \citep{weiler_3d} and implemented in the e3nn library as the basis of our equivariant Unet.  Here we describe how each layer in the UNet has been modified to be equivariant and we explain the details necessary to understand the application to voxelized 3D medical imaging data. 

\subsection{Irreducible Representations}

Typical convolutional neural networks produce scalar-valued output from
scalar-valued features.  A scalar field $f:\mathbb{R}^3 \to \mathbb{R}$ transforms in a very simple way under rotations: the field at the location $x$ after application of a rotation $r$ is given by $f(r^{-1}x)$. However, an equivariant network based purely on scalar fields would have rather minimal representative power, suffering from similar problems as a G-CNN with G-pooling at every layer.  Concretely, such a network would clearly be unable to detect oriented edges.  To enable the learning of expressive functions requires the learning of more general features with a richer law of transformation under rotations.   For example, a vector field assigns a value of $\mathbb{R}^3$ to each point of Euclidean space: one such example is the gradient $\nabla$ of a scalar field.  Such features \emph{are} expressive enough to detect oriented edges; here the orientation is explicit (the orientation of the gradient field).  Under a rotation $r$, a vector field $f$ transforms not as $f(r^{-1}x)$ but as $r f(r^{-1}x)$.

Scalars and vectors are two well-known representations of SO(3), but there are many others. It's worth noting that all finite representations of SO(3) can be broken down into a combination of simpler, indivisible representations known as "irreducible representations", as described in \citep{weiler_3d} and \citep{thomas_tensor_2018}.  
In SO(3), each irrep  is indexed by a positive integer $l = 0,1,2,\dots$  and has dimension $d = 2l+1$  . A major contribution of \cite{weiler_3d} was the formulation and solution of a constraint on kernels between irreps of order $l$ and $l'$, giving rise to a basis of all such kernels: this basis is implemented in the \texttt{e3nn} library.  Networks defined using the operations of  \texttt{e3nn} can have features valued in any irreps.  For our experiments we consider features valued in scalars ($l=0$),  vectors ($l=1$) and rank-2 tensors ($l=2$).

\subsection{Equivariant voxel convolution}

\begin{figure} \center \includegraphics[width=\textwidth]{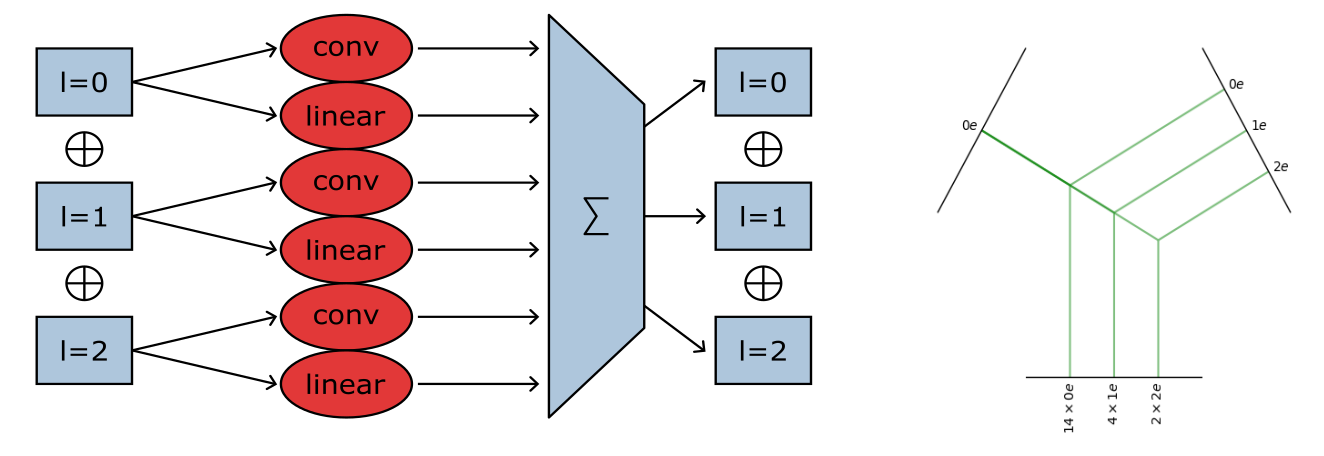}
    \caption{(Left) Our equivariant self-connection convolutional layer for feature extraction: a single irreducible representation is produced by the sum of a convolution on the scalar irreps ($l=0$), a convolution on the vector features ($l=1$) and a convolution on the tensor features ($l=2$), together with a self connection layer (voxel-wise fully connected tensor product between the irreps). (Right) Illustration of the fully connected tensor product in the beginning of our network. The input representations are our scalar image "0e" on the left and the spherical harmonics of $l$ from 0 to 2 on the right, which result in a hidden layer of irreps of scalar, vector and rank-2 tensors.}
     \label{Equi-conv} \end{figure}

Each layer of an equivariant network formulated in \texttt{e3nn} takes as input a direct sum of irreps  and returns a direct sum of irreps (in our case, of orders, $l=$0, 1 or 2) See Fig. \ref{Equi-conv}.

An equivariant convolutional kernel basis is described in \cite{weiler_3d}: the basis functions are given by tensor products $\phi(\lVert x \rVert) Y^l (x / \lVert x \rVert)$.  Here $\phi : \mathbb{R}^+ \to \mathbb{R}$ is an arbitrary continuous radial function describing how the kernel varies as a function of distance from the origin. $Y^l$ is the spherical harmonic of order $l$, determining how the kernel varies within an orbit of SO(3) (a sphere centred on the origin).

The equivariant convolution is described by Equation~\ref{voxel_convolution}.
We introduce the terms in the equation in Table~\ref{tab:inoutconv}.
$F$ and $F'$ are the input and output fields. Each of them have an irrep that determines their dimension and how their transform under rotation.
To calculate the output channel $F'_j(x)$ we sum over the contributions from the input channels. Each contribution is characterized by an input channel index $i$, an input irrep $l_i$, an output channel index $j$ and its irrep $l_j$ and a spherical harmonic order $l$ satisfying the selection rule (Equation~\ref{selection_rule}).

\begin{table}[h]
    \centering
    \begin{tabular}{|c|c|}
    \hline
        Input features & Output features \\
    \hline
        $F_1$ with $l_1 = 0$ & $F'_1$ with $l_1 = 0$ \\
    
        $F_2$ with $l_2 = 0$ & $F'_2$ with $l_2 = 1$ \\
   
        $F_3$ with $l_3 = 1$ & $F'_3$ with $l_3 = 2$ \\
 
        $F_4$ with $l_4 = 2$ & $F'_4$ with $l_4 = 2$ \\
  
                             & $F'_5$ with $l_5 = 2$ \\
    \hline
    \end{tabular}
    \caption{An example to illustrate the notation of Equation~\ref{voxel_convolution} where $F_i$ are the input channels and $F'_j$ are the output channels. Each channel is an irrep-field. In the example shown here, the input field $F_3$ is a vector field (because $l_3=1$), it's therefore a $\mathbb{R}^3 \longrightarrow \mathbb{R}^3$ function. Similarly, $F_4$ is a $\mathbb{R}^3 \longrightarrow \mathbb{R}^5$ function.}
    \label{tab:inoutconv}
\end{table}

 
    \begin{equation}\label{voxel_convolution}
        F'_j (x) =  \sum_{\{{i \times l} \xrightarrow {}j \}} \int da \; F_{i} (x+a) \underset{l_i\times l \xrightarrow{} l_j}{\otimes} Y^{l}(\frac{a}{\|a\|}) \sum_{k} b_k (\|a\|) w(k,i\times l \xrightarrow{} j)
    \end{equation}
    
Each incoming channel $F_i$, and outgoing channel $F'_j$,  has a specified irrep. In this notation, $\{{i \times l} \xrightarrow {}j \}$ denotes a "path" from an input channel $i$ to an output channel $j$ via a spherical harmonics $l$. All irreps $l$ satisfying the \emph{selection rule} of the group SO(3) are the nonzero integer satisfying
    \begin{equation} \label{selection_rule}
        |l_i - l_j| \leq l \leq l_i+l_j 
    \end{equation}
where $l_i$, $l_j$ are the irrep of the input and output channels.   These are the allowed "paths" between the input and output: all the ways in which a feature of irrep $l_i$ can yield a feature of irrep $l_j$ respecting SO(3) equivariance.
The notation $\underset{l_1\times l_2 \xrightarrow{} l_3}{\otimes}$ denotes the tensor product of irrep $l_1$ times irrep $l_2$ reduced into irrep $l_3$: this is unique for the group $SO(3)$ (contrary to $SU(3)$, the special unitary group of degree 3, for instance). Examples are listed in Table~\ref{tab:tp}.

\begin{table}[h]
    \centering
    \begin{tabular}{|c|c|}
    \hline
        $\underset{0\times 0 \xrightarrow{} 0}{\otimes}$ & the normal multiplication of scalars \\
    \hline
        $\underset{0\times 1 \xrightarrow{} 1}{\otimes}$ & scalar times vector, same signature as the gradient $\nabla f$. \\
    \hline
        $\underset{1\times 0 \xrightarrow{} 1}{\otimes}$ & vector time scalar. \\
    \hline
        $\underset{1\times 1 \xrightarrow{} 0}{\otimes}$ & dot product of vectors, same signature as the divergence $\nabla\cdot \vec f$. \\
    \hline
        $\underset{1\times 1 \xrightarrow{} 1}{\otimes}$ & cross product of vectors, same signature as the rotational $\nabla\wedge \vec f$. \\
    \hline
    \end{tabular}
    \caption{Examples of reduced tensor products for the group SO(3). Some of these, in the context of the convolution, can be related to differential operators. But note that the differential operators are local while the convolution is non local.}
    \label{tab:tp}
\end{table}

This calculation is implemented in \texttt{e3nn} by sampling the continuous kernel at the grid points of the voxel grid yielding an ordinary kernel: this kernel is then convolved over the input irreps.  This means that efficient cuda implementations of convolutional layers can be used during training, and that at test time the (rather computationally expensive) tensor product operations can be avoided by precomputing an ordinary CNN from the equivariant network.

Since the radial basis functions all vanish at zero, the convolutional kernels yielded are necessarily zero at the origin: to account for this we also include, at each convolutional layer, a \emph{self connection} layer, which is simply a pointwise weighted tensor product: this can be seen as the equivalent of a convolutional layer with 1 x 1 x 1 kernel. Our feature extractor is then the sum of the convolutional layer and the self connection layer, and it is this layer that we use to replace an ordinary convolution in the Unet architecture.

\subsection{Pooling, upsampling, non-linearities and normalization layers}

 The crucial observation in creating layers compatible with irrep-valued features is that while scalar features can be treated pointwise, as in an ordinary network, the components of vectors and tensors  must be transformed together, rather than treated as tuples of  scalar values.

In line with \cite{weiler_3d} we use \emph{gated} nonlinearities, in which an auxiliary scalar feature calculated from the irreducible feature is used passed through a sigmoid nonlinearity, which is then multiplied with the feature to induce a nonlinear response. 
  
 For the encoding path, we apply ordinary max-pooling to the scalar valued feature components.  For a vector or tensor valued component $v$, we pool by keeping the vector with the greatest $l^2$ norm \citep{cesa2021program}. 
 
 We apply ordinary instance normalization \citep{ulyanov2016instance} to the scalar features. Similarly, to instance-normalize a vector- or tensor-valued feature $v$ we divide by the mean $l^2$
 norm of that feature per instance: $\textrm{norm(v)} := v/{\mathbb{E}( \|v\|)}$

\subsection{Related Work}

To build our proposed architecture we have made heavy use of the e3nn framework for developing equivariant neural networks: however there are other packages for creating equivariant networks available.
The escnn package, \cite{cesa2021program}, in particular provides layers for building E(2)- and E(3)-equivariant networks; indeed, the max-pooling and instance normalization layers are equivalent to those we use in our network.

Previously published rotation-equivariant Unets have been restricted to 2D data and G-CNN layers~\citep{chidester2019enhanced,linmans_sample_2018,pang2020beyond,winkens2018improved}.  A preprint describing a segmentation network based on e3nn filters applied to multiple sclerosis segmentation for the specific use case of 6 six-dimensional diffusion MRI data is available \citep{muller2021rotation}: in this particular setting each voxel carries  three dimensional q-space data, with the network capturing equivariance in both voxel space and q-space. In contrast to the current paper, ordinary (non-equivariant) networks were unable to adequately perform the required segmentation task (lesion segmentation from diffusion data).  This leaves the question open of whether equivariant networks have advantages over plain CNNs in the case of more typical 3D medical data. Here we show that end-to-end equivariant networks are indeed advantageous even when operating on scalar-valued inputs and outputs. 

Other works in the application of equivariant networks to 3D data have focused on classification rather than segmentation, primarily using G-CNNs \citep{andrearczyk2019exploring}.

\section{Methods}

\subsection{Model architectures}

\subsubsection{Irreducible representations}

The design of equivariant architectures offers somewhat more freedom than their non-equivariant counterparts, insofar as we have more degrees of freedom in specifying the feature dimension at each layer: not just how many features, but how many of each irreducible order.  In the majority of experiments in this paper we fixed a ratio 8:4:2 of order 0, 1 and 2 irreps in each layer other than input and output. In the notation of the \texttt{e3nn} library, this combination is denoted \texttt{8x0e~+~4x1e~+~2x2e}, and corresponds to an ordinary feature depth of 30. In order to investigate sensitivity to the feature ratio, we compared to three other combinations of ratios with a similar feature dimension (8:4:2, 30:0:0, 8:8:0, and 4:4:4). This investigation (\ref{irreps_comparison}) suggests that selection of the correct ratio is important, and that our proposed ratio is a good one for the kind of tasks we examined.

\subsubsection{Kernel dimension and radial basis functions}

Aliasing effects mean that if we choose a kernel which is too small, higher spherical harmonics may not contribute (or contribute poorly) to learning.  For this reason, we choose a larger kernel (5x5x5) than often used in segmentation networks.  

In addition to specifying the size of the convolutional kernel we must also specify which and how many radial basis functions are used to parameterize the radial component of the convolutional filters.  We fix five basis functions for each equivariant kernel described in the appendix.

\subsubsection{Reference and equivariant Unet architectures.}

As a reference implementation of Unet we used the nnUnet library \citep{isensee_nnu-net_2021}, with $5^3$ convolutional kernels,  instance normalization, and leaky ReLu activation after each convolutional layer.  The network uses max-pooling layers for downsampling in the encoding path,  trilinear upsampling in the decoding path, and has two convolutional blocks before every max-pooling layer and after every upsampling. The number of features doubles with every max-pooling and halved with every upsampling, in accordance with the usual Unet architecture. 

We mirror this architecture in the equivariant Unet, simply replacing the ordinary convolutions with equivariant convolutions/self-connections (using the ratios of irreps specified above), equivariant instance normalization and gate activation after each convolution.  The network uses equivariant max-pooling layers for downsampling in the encoding path, and trilinear upsampling in the decoding path, and the number of irreps of each order double at each max-pooling and halve with every upsampling. 

\section{Datasets and Experiments}

We carried out a number of experiments to validate the hypothesis that equivariant Unet models are sample efficient, parameter efficient and robust to poses unseen during training.  In all experiments, we used categorical cross entropy as loss function, with an Adam optimizer, a learning rate of $5\text{e-}3$
 and early stopping on the validation loss with a patience of 25 epochs for the brain tumor segmentation task and 150 epochs for the healthy-appearing brain structure segmentation.
 Networks were trained on 128x128x128 voxel patches
and prediction of the test volumes was performed using patch-wise prediction 
with overlapping patches and Gaussian weighting \citep{isensee_nnu-net_2021}.  In all cases, we used the Dice similarity metric to compare the segmentation output of the network to the reference standard.

\subsection{Medical Image Decathlon: Brain Tumor segmentation}

\begin{figure}
        \begin{subfigure}[b]{0.25\textwidth}
\includegraphics[width=\linewidth]{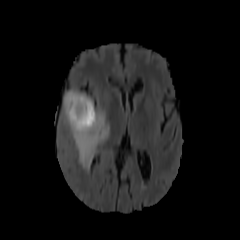}
\caption{FLAIR}
\label{fig:flair}
\end{subfigure}%
\begin{subfigure}[b]{0.25\textwidth}
\includegraphics[width=\linewidth]{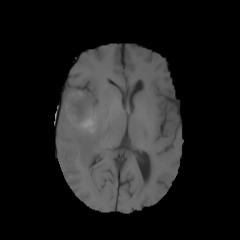}
\caption{T1 weighted}
\label{fig:T1}
\end{subfigure}%
\begin{subfigure}[b]{0.25\textwidth}
\includegraphics[width=\linewidth]{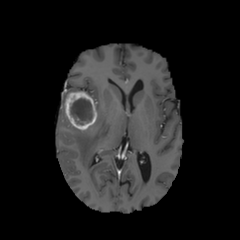}
\caption{T1 postcontrast}
\label{fig:tiger}
\end{subfigure}%
\begin{subfigure}[b]{0.25\textwidth}
\includegraphics[width=\linewidth]{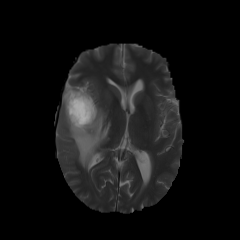}
\caption{T2 weighted}
\label{fig:t2}
\end{subfigure} 
\caption{Four imaging modalities used in the brain tumor segmentation task. The brain tumor can be clearly seen in the top left.}%
\label{fig:brain_contrasts}
\end{figure} 

 484 manually annotated volumes of multimodal imaging data (FLAIR, T1 weighted, T1 weighted postcontrast and T2 weighted imaging) of brain tumor patients were taken from the Medical Segmentation Decathlon \citep{antonelli2021medical} and randomly separated as  340 train,  95 validation and  49 test volumes. The four imaging contrasts are illustrated in Fig.~\ref{fig:brain_contrasts}. We trained both an equivariant Unet and an ordinary Unet, each with three downsampling/upsampling layers, for the task of segmenting the three subcompartments of the brain tumor.  The basic Unet had a feature depth of 30 in the convolutions of the top layer, with the equivariant network having an equivalent depth of 30 features.  The Unet was trained both with and without rotational data augmentation (rotation through an angle $\in (0,360)$ with bspline interpolation), on both the full training set and also subsets of the training set (number of training samples was $2^n$ for $n$ between 1 and 9, inclusive).  With this we aim to study the sample efficiency of the two architectures. The data-augmented Unet needed longer training with a patience of 100 epochs.
 
 We do not expect orientation cues to be helpful in segmenting brain tumors (which are largely isotropic) and therefore expect that both the ordinary and equivariant Unet will maintain performance under rotation of the input volume, and that data augmentation will be primarily useful where amounts of training data are small.  
 
 \subsection{Mindboggle101 dataset: Healthy appearing brain structure segmentation}

\begin{figure}[!htb]
\centering
\subfloat[\centering Sagittal view: cerebellum (red), hippocampus(green), lateral ventricle (dark blue), caudate (yellow), putamen (cyan), pallidum (pink) ]{{\includegraphics[width=6cm]{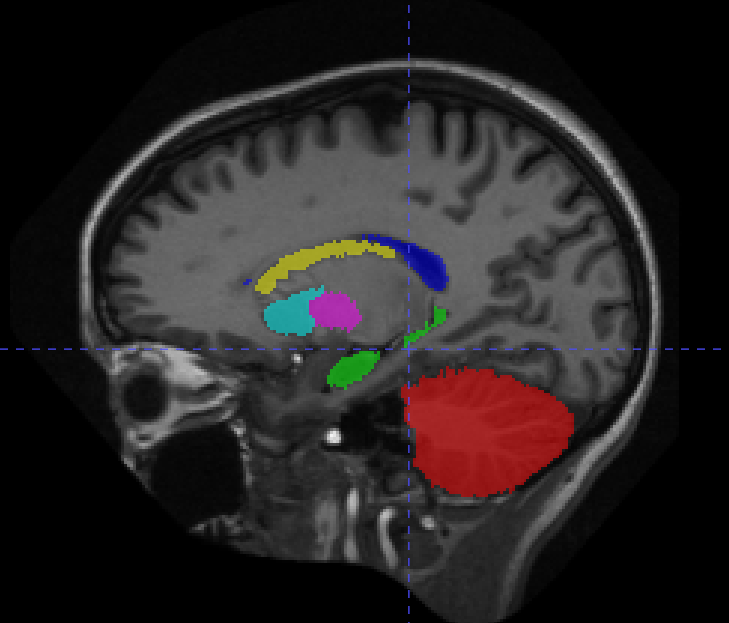} }}%
\qquad
\subfloat[\centering Coronal view: cerebellum (red), brain stem (peach)]{{\includegraphics[width=5cm]{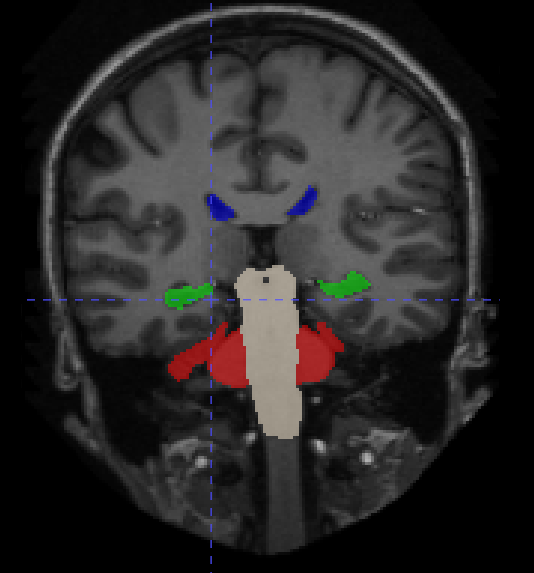} }}%
\caption{Two cross sections showing the seven brain structures chosen for the healthy-appearing brain structure segmentation task}%
\label{fig:mindboggle_labels}%
\end{figure}

From the 20 manually annotated volumes of the Mindboggle101 dataset \citep{klein2012101}, we selected 7 volumes for training, 3 volumes for validation and 10 volumes for testing. The Mindboggle101 labelling contains a very large set of labels, including both cortical regions and subcortical structures. We defined the following subset of structures as target volumes for segmentation: cerebellum, hippocampus, lateral ventricles, caudate, putamen, pallidum and  brain stem. These structures are shown in figure~\ref{fig:mindboggle_labels}.  Some of these structures (ventricles, cerebellum) can be easily identified by intensity or local texture, while others (caudate, putamen, pallidum) are difficult to distinguish except by spatial cues.  As above, an equivariant Unet and an ordinary Unet, each with three downsampling/upsampling layers, were trained to segment these structures: again both networks had an equivalent feature depth (30 features).  The ordinary Unet was trained without data augmentation and with full rotational data augmentation (rotation through an angle $\in (0,360)$ in either the axial, saggital or coronal plane, with bspline interpolation). The ordinary Unet was trained a third time with a data augmentation scheme closer to that seen in usual practice (rotation through an angle $\in (-20,20)$ in either the axial, sagittal or coronal plane, with bspline interpolation).  Once trained, these models were then applied to the testing set rotated through various angles $\in (0,180)$, to test the sensitivity of the various models to variations in pose.  To provide a comparison between e3nn-style steerable filters and group convolutions we also trained trained unet-style models with layers derived from the gconv package.

Finally, we trained variants of the Unet and equivariant Unet with increasing model capacity, as measured by the equivalent feature depth in the top layer, and applied these models to the unrotated test samples to assess the parameter efficiency of the two architectures.  Here no data augmentation was employed.

\section{Results} 

\subsection{Brain Tumor segmentation}

\begin{table}[!htb]
\centering

\begin{tabular}{|l|c|c|c|}
\hline
Model & Enhancing Tumor & Tumor Core & Whole Tumor  \\
\hline
e3nn   & $0.85\pm 0.12$ & $0.88 \pm 0.07$ & $0.92 \pm 0.06$  \\
nnUnet & $0.78 \pm 0.19$ & $0.84 \pm 0.08$ & $0.90 \pm 0.06$ \\
nnUnet (da) & $0.76 \pm 0.02$ & $0.83 \pm 0.07$ & $0.90 \pm 0.06$ \\
\hline
\end{tabular}

\caption{Dice score on the test set for the brain tumor segmentation task. nnUnet (da) denotes the non-equivariant reference network trained with data augmentation.}\label{tab1}

\end{table}

\begin{figure} 
\center \includegraphics[scale=0.8]{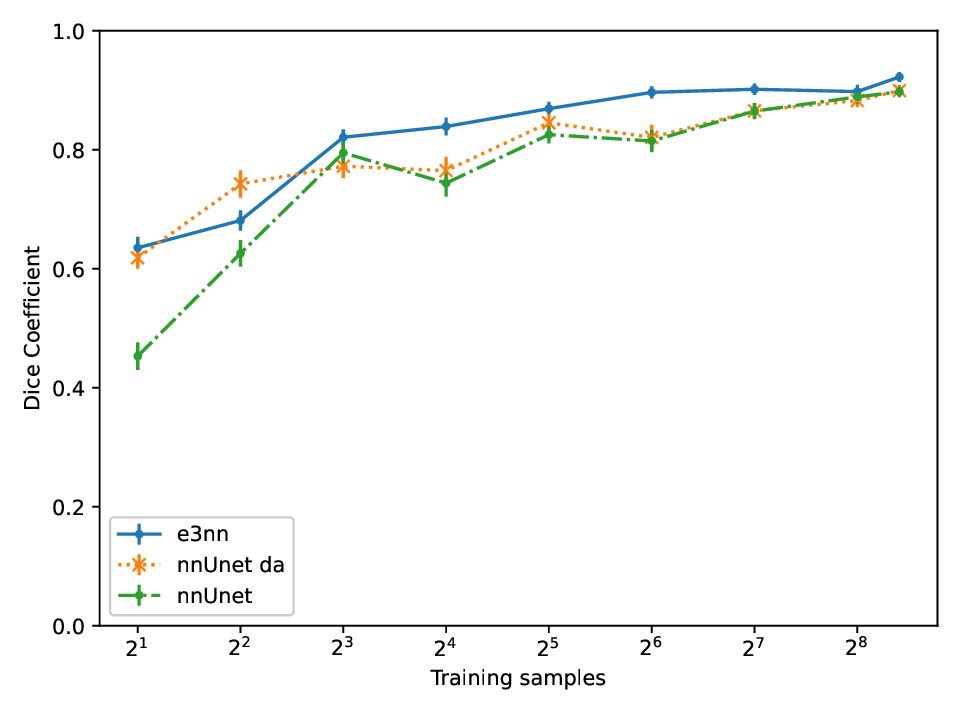}
    \caption{Dice score on the test set for the brain tumor segmentation task as a function of the number of volumes used for training.} \label{perf_vs_samples} \end{figure}
Table~\ref{tab1}  shows the performance of the equivariant Unet (e3nn) versus a reference Unet (nnUnet) with and without data augmentation, over the 49 testing examples. In Figure~\ref{perf_vs_samples} we show performance on the testing set for networks trained on subsets of the training volumes averaging over all compartments.  The gap in performance between the  equivariant and reference networks is largest where data is scarce, and as expected this is also where data augmentation has the largest effect on the performance of the ordinary Unet.

\subsubsection{Memory requirements and computation time}
We trained both models on a workstation with a NVIDIA RTX A6000 graphics card. During training the reference model took an average of 400 s and 7.8 GB of memory per epoch. Our equivariant model took 518 s and 13.3 GB of memory. 

\subsection{Healthy-appearing brain-structure segmentation}

\subsubsection{Effect of input irreps ratios} \label{irreps_comparison}
\begin{table}[!h]
\begin{tabular}{|l|c|c|c|c|} 
\hline
\bf{Brain structure} & \bf{8:4:2 (30)} & \bf{30:0:0 (30)} & \bf{8:8:0 (32)} & \bf{4:4:4 (36)}  \\ \hline

cerebellum & $\bm{0.952 \pm 0.001}$ & $0.938 \pm 0.005$ & $0.949 \pm 0.006$ & $0.938 \pm 0.004$ \\
hippocampus & $\bm{0.834 \pm 0.009}$ & $0.801 \pm 0.009$ & $0.82 \pm 0.01$ & $0.79 \pm 0.01$ \\
lateral ventricle & $0.89 \pm 0.01$ & $0.89 \pm 0.01$ & $\bm{0.90 \pm 0.01}$ & $0.87 \pm 0.02$ \\
caudate & $\bm{0.888 \pm 0.003}$ & $0.878 \pm 0.006$ & $0.887 \pm 0.004$ & $0.86 \pm 0.01$ \\
putamen & $\bm{0.889 \pm 0.003}$ & $0.863 \pm 0.007$ & $0.888 \pm 0.005$ & $0.881 \pm 0.004$ \\
pallidum & \bm{$0.854 \pm 0.006}$ & $0.830 \pm 0.005$ & $0.849 \pm 0.005$ & $0.74 \pm 0.02$ \\
brain stem & $0.912 \pm 0.002$ & $0.899 \pm 0.004$ & $\bm{0.917 \pm 0.002}$ & $0.880 \pm 0.003$ \\ \hline

\end{tabular}

\caption{Dice score on the healthy-appearing brain structure segmentation task for different irrep ratios (scalars to vectors to rank-2 tensors). The number in parenthesis after each ratio denotes the equivalent scalar feature depth for that ratio. Entries in bold denote the best performing ratio for each structure. The 8:4:2 ratio, which we use for all other experiments in this paper, is the best performing in 5 of the structures.}  
\label{table:irreps} 
\end{table} 

We have chosen four different ratios of scalars to vectors to rank-2 tensors to understand the effect of this hyperparameter on the performance on the segmentation network. We chose four ratios which have as close a feature dimension as possible. These dimensions and the dice scores are listed in table \ref{table:irreps}. The results indicate a clear benefit from having vector irreps rather than just scalar irreps.  On the other hand, the performance of the network with a larger ratio of  rank-2 tensors substantially underperforms, while a model with no rank-2 features but an increased number of vector features performs almost as well as the best performing model (the ratio of 8:4:2, used for all the other experiments in this paper.)  While this experiment validates the ratio used in our experiments, this does not rule out the possibility that other ratios may perform better in different segmentation tasks.

\subsubsection{Performance on the test set}

\begin{figure}[!htb]
    \begin{center}
    \includegraphics[scale=1]{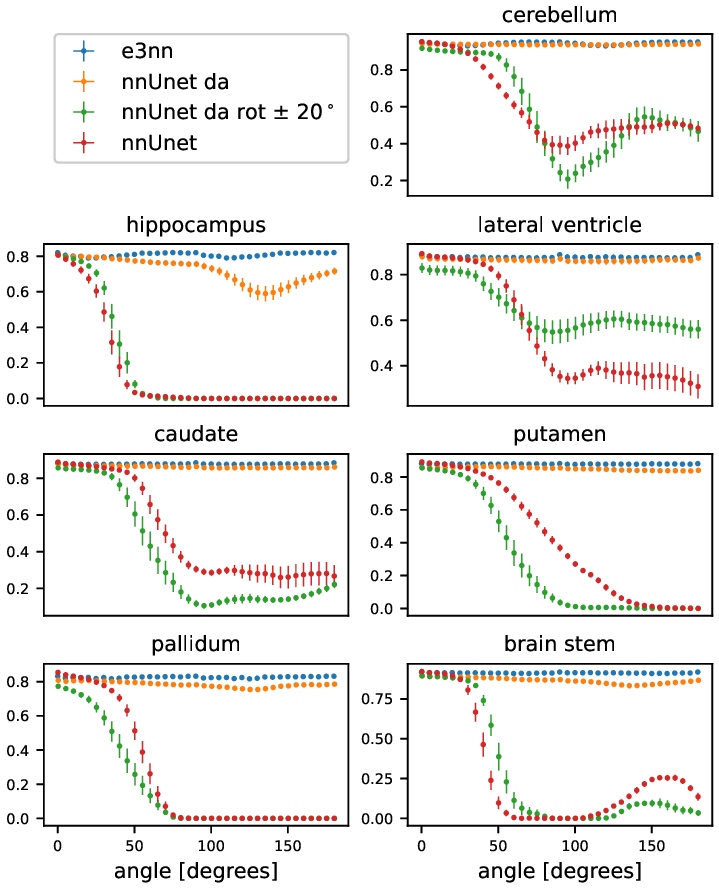}
    \caption{Dice score on the test set vs rotation angle in the axial plane on the brain structure segmentation task.}
    \label{fig:mindboggle_fig} 
    \end{center}
\end{figure}

In Figure~\ref{fig:mindboggle_fig} we show the results of the segmentation of the various brain structures of the Mindboggle 101 dataset. For a non-rotated version of the test set, our proposed network (e3nn) and the reference network without rotation augmentation (nnUnet) performed similarly well in segmenting all structures except the pallidum and putamen, where our proposed network showed a higher performance. Moderate data augmentation of angles less than 20$^\circ$ (nnUnet da rot 20) had a negative effect on performance in some structures when compared to the reference network. The nnUnet trained with full rotational data augmentation performed on par with the proposed network but underperformed on segmentation of the hippocampus and pallidum.

\subsubsection{Performance on rotated inputs from the test set}

When tested on rotated versions of the testing volumes the non-equivariant reference network's performance smoothly declines even for angles $<$ 20$^\circ$.  The equivariant network, as 
expected, is not affected by rotations: small fluctuations in Dice coefficient can be accounted for by interpolation artifacts in the input image. Data augmentation up to 20$^\circ$ does not seem to improve performance very much compared to the reference network. In fact, this network performs worse than the reference network in three of the seven structures. The reference network with full data augmentation showed good rotational equivariance in all structures except for the hippocampus. 
We only show the results of the rotation experiment when rotated in the axial plane but similar results were obtained when rotating in the coronal and sagittal plane (see appendix).

\subsubsection{Performance as a function of number of parameters}
\begin{figure}[!htb] 
    \begin{center}
    \includegraphics[scale=0.9]{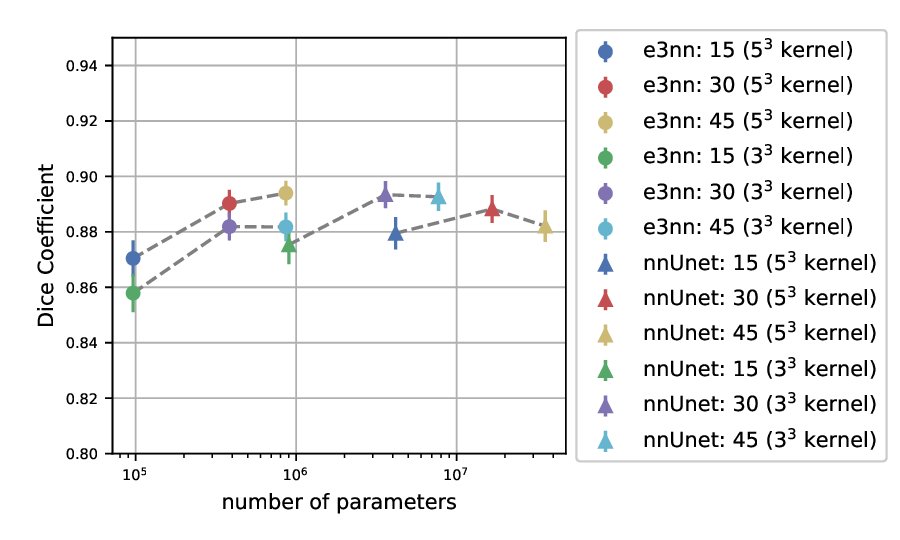}
    \caption{Dice score on the brain structure segmentation task as function of number of trainable parameters. The number next to each model specifies the number of top level features. The error bars display the uncertainty of the mean estimator of the dice score across both the 7 structures and 10 images in the test set. Dashed lines were added for readability, grouping the two kernel sizes for each model.}
    \label{fig:mindboggle_perf}
    \end{center}
\end{figure}
We trained the SO(3) equivariant model and non-equivariant reference Unet with different numbers of input features. We used the dimension of input $dim=2l+1$ of each equivariant model's top features to set the value of the reference Unet top features. Figure~\ref{fig:mindboggle_perf} shows the dice score vs total number of trainable parameters for various numbers of top level features of both models. The rotation-equivariant model has fewer parameters than any of the reference Unet implementations. We also included versions of the both models trained on $3^3$ kernels, which is the kernel size generally-used. Our proposed model with  $5^3$ kernels outperforms the proposed model with $3^3$ kernels, even though they have the same number of parameters.

\subsubsection{Comparison with a group-convolutional neural networks}

We defined a group-convolutional Unet using the gconv python package \citep{kuipers2023regular} for the purpose of a performance comparison with e3nn-style steerable filters. Group convolutions capture approximate invariance to SO(3) symmetry by applying each kernel multiple times in a variety of rotated forms, according to a subgroup of the full SO(3) symmetry group, and storing those rotated versions in an extra feature dimension.  This means that gconv convolutions induce an additional overhead in terms of both computation and GPU memory. For this reason, we were only able to explore the performance of \replaced{group}{graph} convolutions using a group size of 4 \added{which corresponds to the group of 180$^\circ$ rotations.} \added{Practically, the input to the G-CNN-based Unet takes an extra parameter specifying the group size in addition to the number of input features (equivalent to channels in a regular Unet).}  We compare two variants of this GCNN-based Unet: one with \added {group size of 4 and }8 input features (G-CNN 8), which has a similar memory footprint to a regular Unet with 30 \replaced{input channels}{top-level features}, and one with \added{group size of 4 and }30 input features (G-CNN 30), the same number \added{of channels} as the regular Unet, \added{but }at a cost of a substantially larger memory footprint.

\begin{table}[!h]
\centering
\begin{tabular}{|l|c|} \hline
\bf{model} & \bf{dice score} \\ \hline

e3nn & $0.88 \pm 0.01$ \\
nnUnet & $0.89 \pm 0.01$ \\
nnUnet da & $0.86 \pm 0.01$ \\
nnUnet da rot $\pm$ 20$^\circ$ & $0.84 \pm 0.01$ \\
G-CNN 8 & $0.80 \pm 0.01$\\
G-CNN 30 & $0.82 \pm 0.01$ \\ \hline

\end{tabular}
\caption{Dice score on the healthy-appearing brain structure segmentation task over all of the structures on the unrotated test set.}  
\label{table:gnn} 
\end{table} 

We note that for the same training conditions, the group-equivariant Unets perform worse than both the baseline Unet and our proposed architecture. It is, however notable that despite the relatively poor performance both G-CNNs were robust to rotations, showing good approximate equivariance, despite the size of the group used to encode the equivariance: results are shown in the appendix for test set rotations angles up to 20$^\circ$.

\section{Conclusion} 

In this paper we have presented a variant of the Unet architecture
designed to be used for any volumetric segmentation task in which the predicted label set is invariant to Euclidean rotations. The network can be used as a
drop-in replacement for a regular 3D Unet without prior knowledge of the
mathematics behind the equivariant convolutions, with equivalent or better performance on in-sample data, 
no need to train using (potentially computationally expensive) data augmentation and SO(3) or O(3) equivariance "for free". \added{The user simply needs to specify the input and output irreps (scalars, vectors, etc). For a typical segmentation task the input and output irreps are simply scalars, but our network can output velocity fields, and any other higher-rank representation as well.} This equivariance mathematically
guarantees good performance on data with orientations not seen during training.  This effect is dramatically superior to usual data-augmentation strategies.  As our experiments show, a small amount of augmentation may have no effect, a mild positive effect, or a mild negative effect: this may be due to competing effects: the addition of rotated examples to the training pool increases the total amount of information available to the classifier but may also introduce erroneous training examples owing to interpolation artifacts in the images, the labels, or both.  This may explain the reduced performance of the baseline Unet with data augmentation in the case of Brain Tumor segmentation.  

Our experiments also support the hypothesis that an equivariant
network can learn from fewer training samples compared to a reference network, performs better in
segmentation of oriented structures and has much fewer parameters than an equivalent non-equivariant model.
While we have focused on a single architecture in this paper,the types and number of top level features, number of downsample operations, kernel size and normalization can be easily customized in our library.  Also customizable is the kind of symmetry enforced by the network.  The experiments focused in this paper on SO(3) rather than O(3) equivariance (enforcing equivariance also to inversions) but our implementation has the option to easily create models with O(3) equivariance as well.

For the vast majority of medical image segmentation and classification tasks, we anticipate that using rotation equivariant convolution layers will lead to similar or improved performance, depending on the importance of relative spatial cues to the task and the degree of pose variability in the test population.  In this case, equivariant networks could be used as a drop-in replacement for ordinary networks:  in the worst, case, if there is zero pose variation in the test population, using an equivariant network may result in training time overhead with zero benefit at test time. It is possible to imagine tasks where pose is correlated with the task labels: for example, head angulation in MRI and CT may be correlated with deficit in acute stroke imaging.   In such a case, using equivariant kernels may lead to worse performance: however, we would argue that such correlations are spurious in nature.  In general, before using an equivariant network for a given task the modeller should reflect on whether equivariance is relevant for the task. An obvious example of a case where equivariance would directly hurt performance is the prediction of acquisition direction (axial, sagittal, coronal) from raw data.  A more realistic example would be the segmentation of left/right brain hemisphere: here an O3 equivariant network, equivariant to inversions, would need to rely on anatomical rather than spatial cues to distinguish the two hemispheres, a much more challenging task.

\subsubsection*{Limitations}
We consider in this paper two  kinds of equivariant feature extractors: SO(3)-equivariant kernels based on one author's previous work in \cite{weiler_3d} and group-equivariant CNNS.  
We have limited ourselves to a group size of 4 for the G-CNN comparisons, because memory restrictions do not allow us to train with the same size input patches, and higher group sizes would need a reduced input-feature size. The G-CNN models that we have trained exhibited some equivariance but their performance was much lower than our other models.

We have limited ourselves to data with publicly available images \emph{and} labels, in order to maximize reproducibility: in particular, the experiments on the Mindboggle experiments do not have sufficient statistical power to show a significant difference between the methods examined.  Nonetheless, we believe the effects of equivariance on his publicly available data are compelling on their own and are confident that a reproduction on a much larger dataset (trained and evaluated on, for example, Freesurfer outputs) would show similar results, albeit in a somewhat less reproducible fashion. 

We used a fixed learning rate and training strategy for each network:  thorough hyperparameter tuning would almost certainly improve the performance of each network presented here. Nonetheless, we believe the experiments here are sufficient to support our claims: that equivariant Unets can be used as a drop-in replacement for more commonly used Unets without loss of performance and with substantial advantages in data and parameter efficiency. 

\subsection*{Remarks and Future Work}

Code to build equivariant segmentation networks based on \texttt{e3nn} for other tasks is available at~\url{http://github.com/SCAN-NRAD/e3nn_Unet}. This library supports not just scalar inputs and outputs, but also inputs and outputs valued in any irreducible representation.  In the future it will be interesting to examine possibility of using odd-order scalar network outputs to segment structures with bilateral symmetry using E(3) equivariant networks, and to investigate whether equivariant networks with vector valued outputs are more robust than ordinary convolutional networks in, for example, the task of finding diffeomorphic deformation fields ~\citep{balakrishnan2019voxelmorph}.

\acks{This work was supported by Spark Grant CRSK-3\_195801 and by the Research Fund of the Center for Artificial Intelligence in Medicine, University of Bern, for 2022-23.}


\ethics{The work follows appropriate ethical standards in conducting research and writing the manuscript, following all applicable laws and regulations regarding treatment of animals or human subjects.}

\coi{We declare we don't have conflicts of interest.}

\bibliography{refs}

\section{Appendix}
\subsection*{Radial Basis Functions}
Since equivariance to rotation implies factorization of the kernel into a radial and angular component, the radial component has to be parameterized. These functions are chosen to be smooth and go to zero at the cutoff radius. To enable learning of parameters, we characterise the radial function as a sum of smooth basis elements.  The equation is given by:
\begin{equation}\label{smooth_finite}
8.433573~sus(x + 1) sus(1 - x)
\end{equation}
with $sus$ (soft unit step) defined as follows:
\begin{equation*}
sus(x) = \left\{
        \begin{array}{ll}
            e^{-1/x} & \quad x > 0 \\
            0 & \quad x \leq 0
        \end{array}
    \right.
\end{equation*}
Equation~\ref{smooth_finite} is a $C^\infty$ function and is strictly zero for $x$ outside the interval $[-1, 1]$. The prefactor $8.433573$ ensures proper normalization of the neural network and was obtained empirically.


\subsection*{Rotation results up to 20$^\circ$ around the three different planes}
In the following figures, we show the dice performance of the rotation-equivariant model G-CNN models and reference network with no data-augmentation, with data augmentation up to 20$^\circ$ and full data augmentation on the test set rotated through angles from 0$^\circ$ to 20$^\circ$.
\begin{figure}
    \begin{center}
    \includegraphics[scale=0.9]{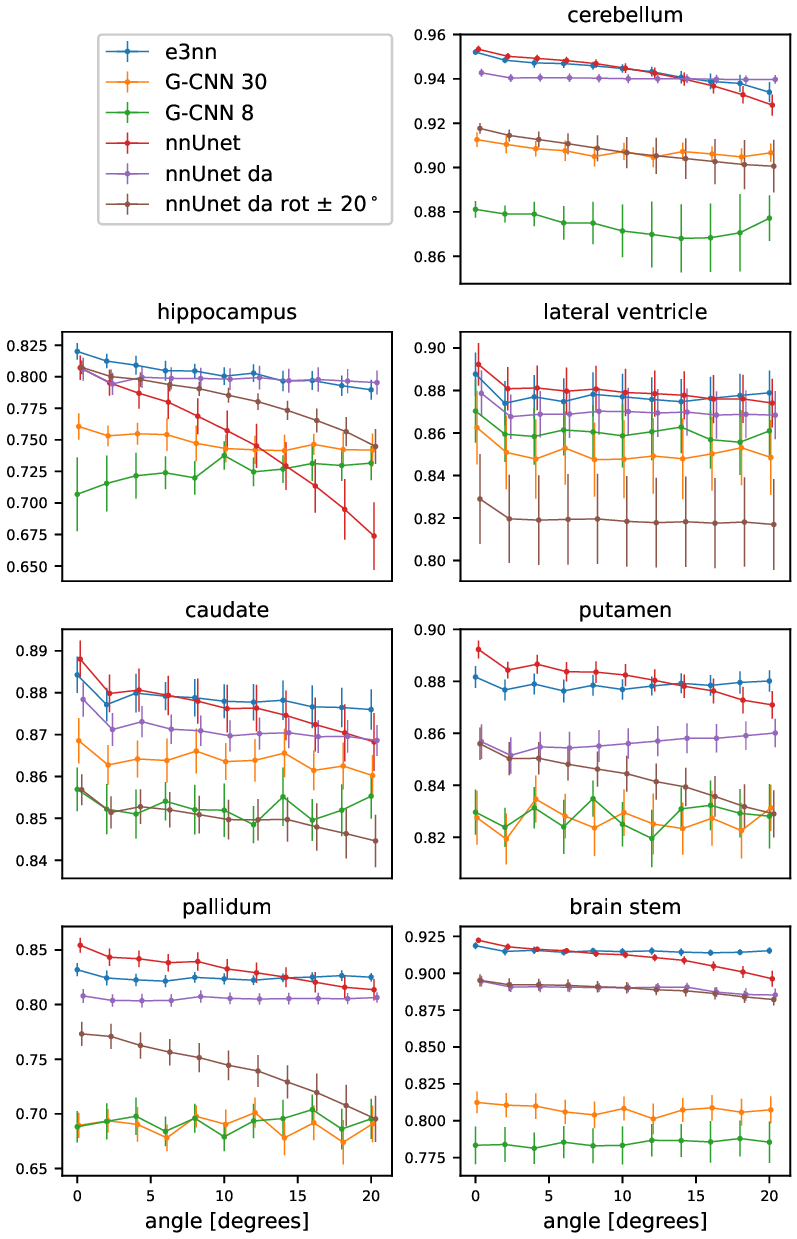}
    \caption{Dice score on the test set vs rotation angle in the saggital plane for seven
    brain structures. The error bars display the uncertainty on the mean estimator of the average dice score across the 10 samples in the test set.}
     \label{mindboggle_rot_20_sag} 
     \end{center}
\end{figure}

\begin{figure} 
    \begin{center}
    \includegraphics[scale=0.9]{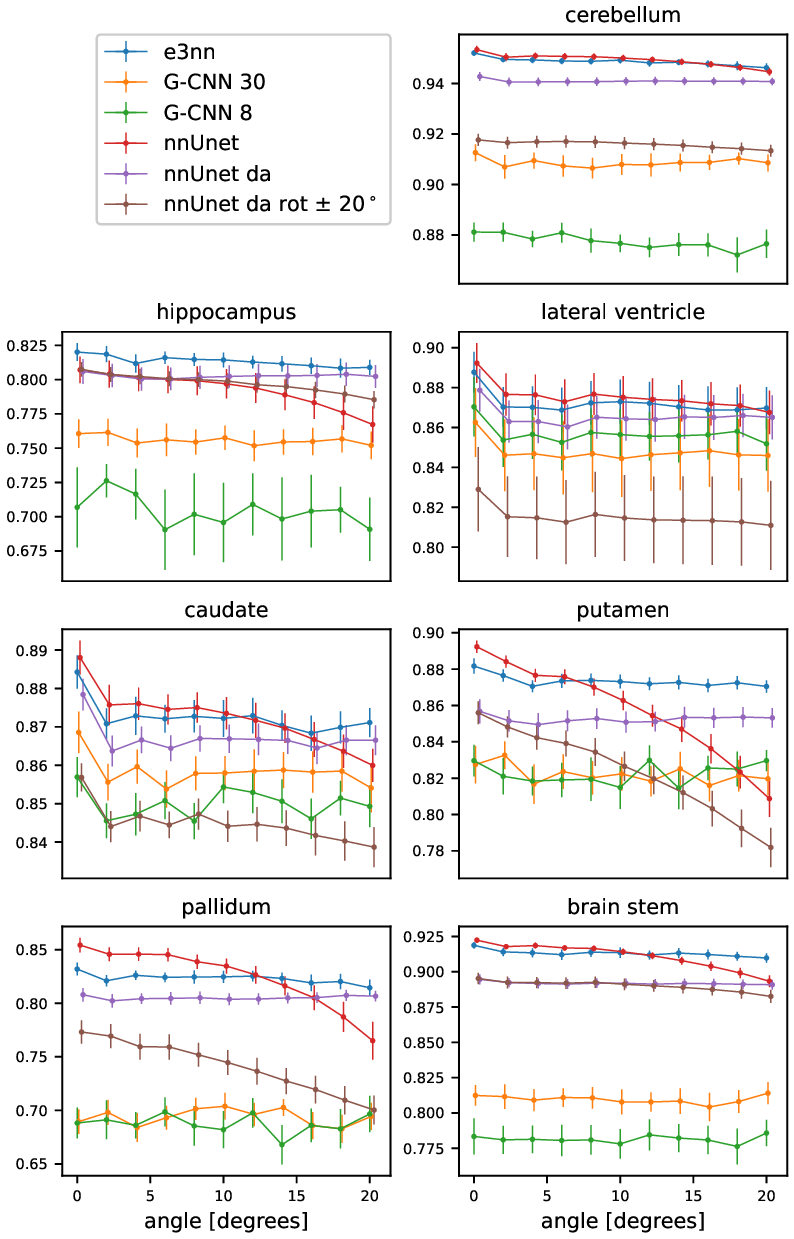}
    \caption{Dice score on the test set vs rotation angle in the coronal plane for seven
    brain structures.}
     \label{mindboggle_rot_20_cor} 
     \end{center}
\end{figure}

\begin{figure}
    \begin{center}
    \includegraphics[scale=0.9]{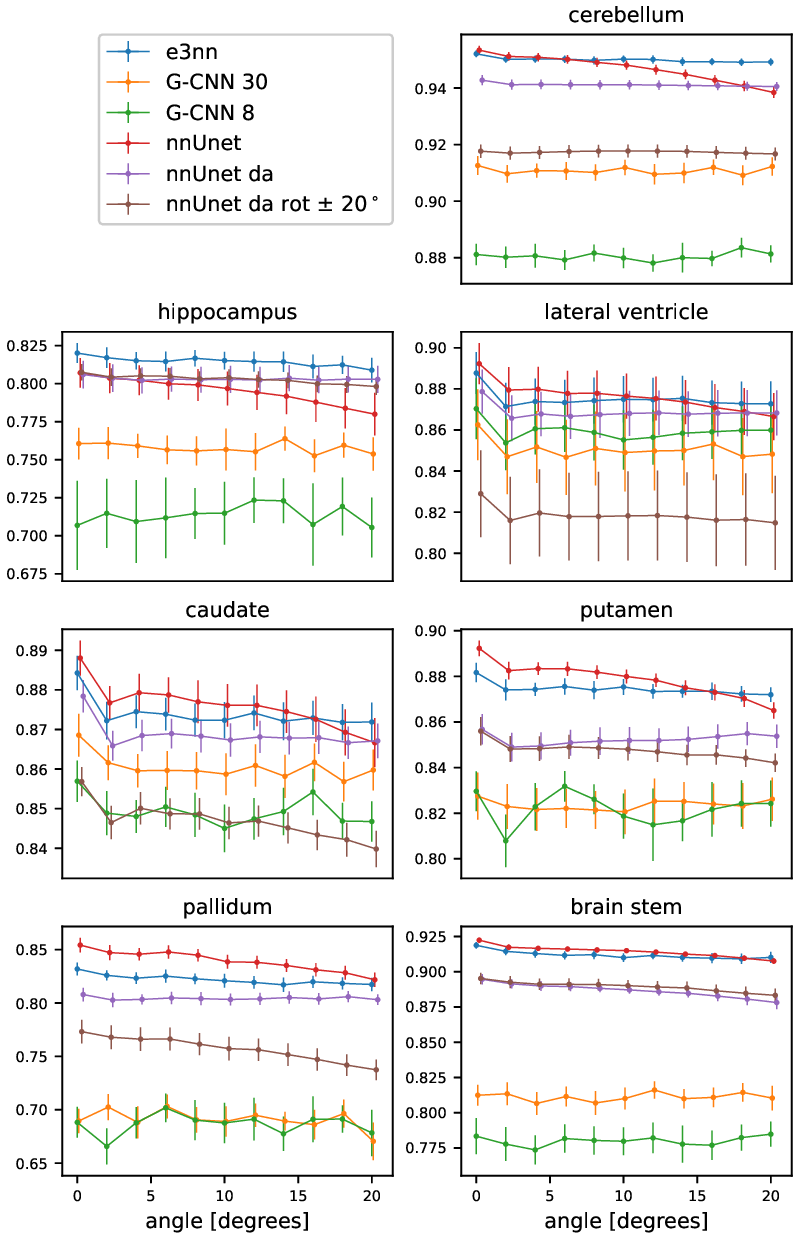}
    \caption{Dice score on the test set vs rotation angle in the axial plane for seven
    brain structures.}
     \label{mindboggle_rot_20_axi} 
     \end{center}
\end{figure}

\end{document}